\documentstyle[12pt]{article}
\newcommand{\ul}{\underline}
\begin{document}
\begin{flushright}
THEF-NYM-94.03
\end{flushright}

\vspace{2\baselineskip}

\begin{center}
{\large\bf THE ONE-BOSON-EXCHANGE \\[2mm]
 POTENTIAL MODEL APPROACH} \\[1.5cm]
J.J.\ DE SWART, P.M.M. MAESSEN and TH.A. RIJKEN \\[2mm]
{\it Institute for Theoretical Physics, University of Nijmegen  \\
Nijmegen, The Netherlands}
\end{center}

\vspace{0.5cm}

\begin{center}
ABSTRACT 
\end{center}

A review is given of the present situation in $Y\!N$ scattering. Special
attention is given to the handling of SU(3) in the various meson exchanges.
The importance of the almost always ignored contribution of the Pomeron is
reiterated.

\vfill

\begin{center}
Invited talk given by J.J. de Swart at the U.S./Japan Seminar on the $Y\!N$-%
Interaction, held at Maui, HA, USA, 25--28 October 1993.
\end{center}

\newpage
\pagenumbering{arabic}
\section{Introduction}         \label{I}
The experimental data on the hyperon-nucleon ($\Lambda N,\ \Sigma N$, and
$\Xi N$) and the hyperon-hyperon $(\Lambda\Lambda,\ \Lambda\Sigma,\
\Sigma\Sigma)$ interactions are very scarce and have moreover large errors.
To give a satisfactory description~\cite{Sw71,Sw88,div}
of these data one needs a {\bf large}
theoretical input. This input is then not allowed to have too many free
parameters, because the scarce data do not allow us to determine reliably
too many parameters. The strategy is therefore to start with a known
description of the $N\!N$-data~\cite{Na75,Na78,Na79}, 
then apply SU(3) flavor symmetry to this
$N\!N$-model in order to obtain this way an $Y\!N$-model~\cite{Na77,Ma89,Na79}. 
Such an approach
can only be successful, when the relevant $N\!N$-model is already consistent
with SU(3). Many models of the $N\!N$-interaction are not suitable
for such an SU(3) generalization. An example of this is the Paris 
model~\cite{La80}. To calculate the two-pion-exchange potential for
$N\!N$ in this model one needs phenomenological input from $\pi N$-scattering.
For the $YN$-interactions one needs the analogous results 
from $\pi\Lambda$ and $\pi\Sigma$ scattering. Such results are not available.

The possibility of SU(3) generalization has in $N\!N$-models implications
for the exchanged mesons. In $N\!N$-potentials one needs not only the
pion-exchange potential, but one needs also the potentials due to the exchange
of the other non-strange members $\eta$ and $\eta^{\prime}$
of the same pseudoscalar octet. Next to the $\rho$ and $\omega$
exchange potentials, one also needs to include the $\phi$ exchange
potential. An $N\!N$-potential model is not suitable for SU(3) generalization,
when it contains only $2\pi$-exchange, because it should also contain
from the outset $\pi\eta$, $\pi\eta^{\prime}$, $\eta\eta$, etc.\ 
exchange potentials. It is clear that not every $N\!N$-model is 
suitable for generalization to $Y\!N$ and $YY$. The Nijmegen potential
models~\cite{Na75,Na78,Na79,Rij93} have always been constructed with this 
generalization to $Y\!N$ and $YY$ in mind.

\section{The experimental data}  \label{II}
It is interesting to compare the description of the $Y\!N$-interaction
directly with our description of the $N\!N$-interaction. In the Nijmegen
partial wave analyses~\cite{Nijm93}
Nijm~PWA93 of the $N\!N$-scattering data with
$T_{\rm lab} < 350$ MeV we have in $pp$-scattering 1787 datapoints and we 
use 21 model parameters. 

In $np$-scattering we have 2514 datapoints and for a good description
of these data in our PWA we use 19 {\bf extra} model parameters. In the model
we have roughly speaking about 100 datapoints per parameter. This allows
for a good determination of these parameters in $N\!N$. 

\begin{figure}[t]
\vspace{8cm}
\includegraphics{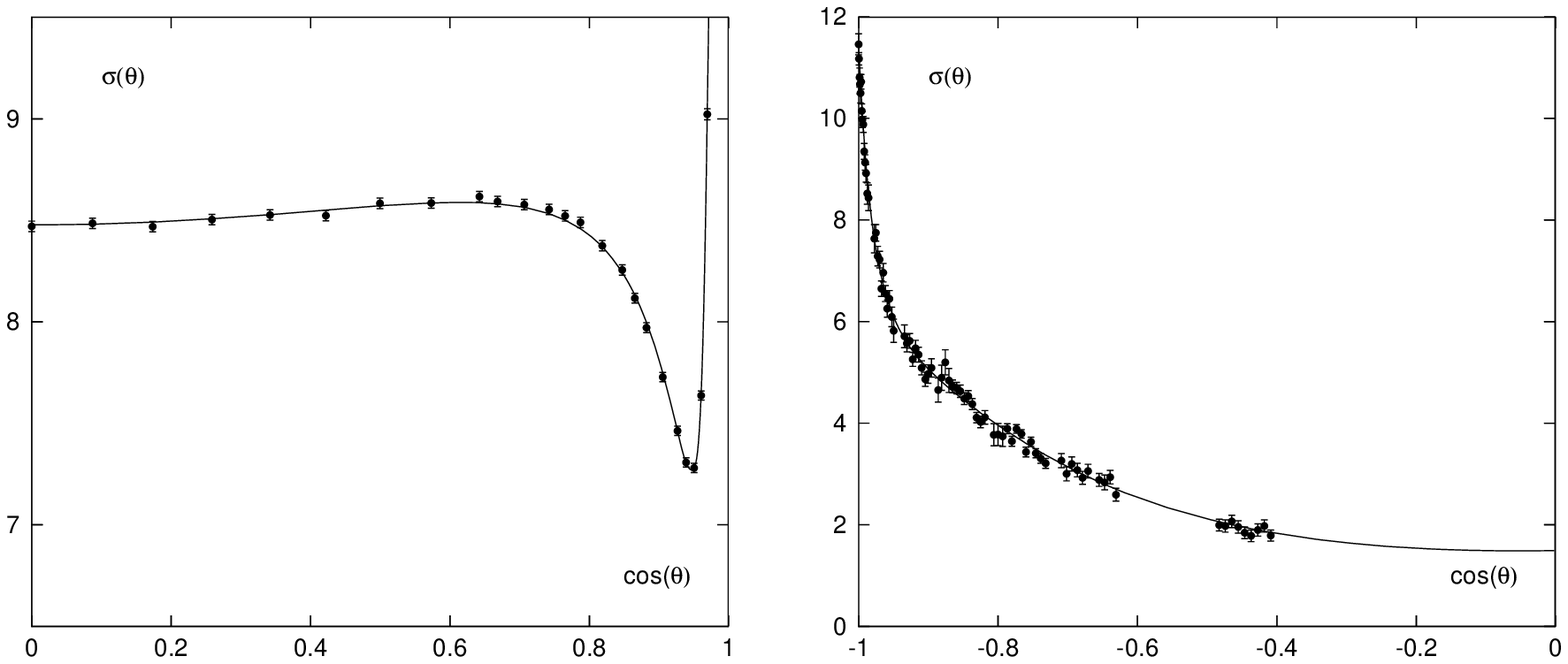}
\caption{(a) $pp$ differential cross section at 50.06 MeV~\protect\cite{Be86}. 
 The 24 datapoints contribute $\chi^{2}=12.8$ in the Nijm~PWA93.\protect\\
 (b) $np$ backwards differential cross section at 344.3 
 MeV~\protect\cite{Bo78}. 
 The 80 datapoints contribute $\chi^{2}=74.53$ to the Nijm~PWA93 and 
 have the normalization 1.035 $\pm$ .005.}
 \label{fig.1}
\end{figure}

In Figure~\ref{fig.1} we show $pp$ and $np$
differential cross sections. 
Shown are the datapoints with their error bars
and the fit of the Nijm~PWA93. We show this to indicate the quality difference
between these data sets and the $Y\!N$-data sets. 

\begin{figure}[bt]
\vspace{7.5cm}
\includegraphics{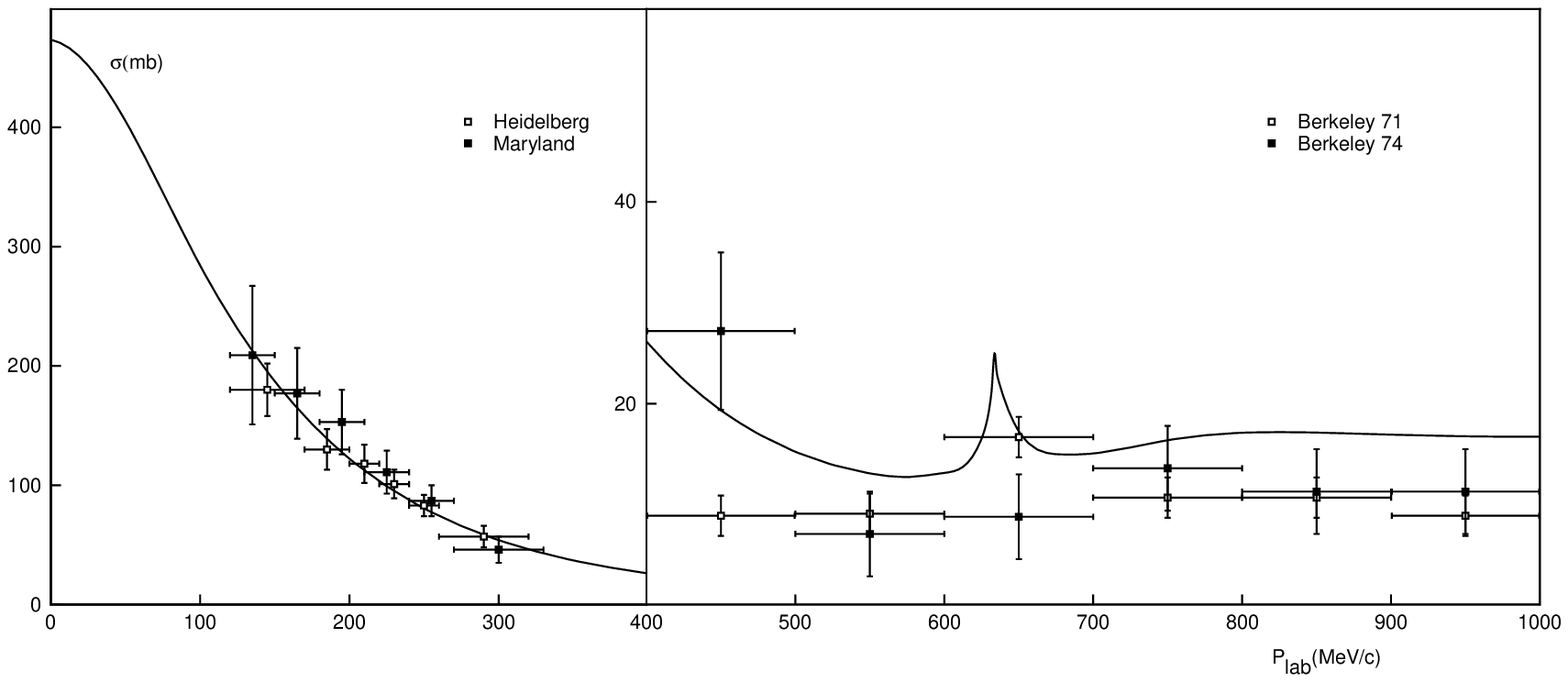}
\caption{Cross sections for elastic $\Lambda p$ scattering. 
 The data are taken from \protect\cite{Al68,Se68,Ka71,Ha77}.}
\label{fig.3}
\end{figure}

For the $Y\!N$-channels it has been customary to use a set of 35 selected 
datapoints~\cite{Na73a}. This is essentially the only scattering information
available about the low energy $Y\!N$-interaction. The data were obtained
from an experiment of slopping $K^{-}$-mesons in the 81 cm Saclay hydrogen
bubble chanber at CERN.  There are a few
extra scattering data available, but these extra data do not really carry 
extra information.
Important to note is, that these data stem from prior to
1971. Finally one has the hyperfragment data~\cite{Gi}, 
which supply some insight in the $Y\!N$-interactions. 

This selected data set of $Y\!N$-scattering data 
is described below. The predictions in the figures
correspond to the unpublished Nijmegen SCW-model~\cite{Mae}, which fits these
35 data with the $\chi^{2}=16.9$.

For elastic $\Lambda p$ scattering (see Figure~\ref{fig.3}) there exist
12 datapoints in the momentum range 120 MeV/c $< p_{\rm lab} < 330$ MeV/c,
which corresponds to the kinetic energy in the laboratory system
6.5 MeV $< T_{\rm L} < 50$ MeV. From these data 6 come from the
Rehovoth-Heidelberg group~\cite{Al68} and 6 come from the Maryland
group~\cite{Se68}.

\begin{figure}[tb]
\vspace{6cm}
\includegraphics{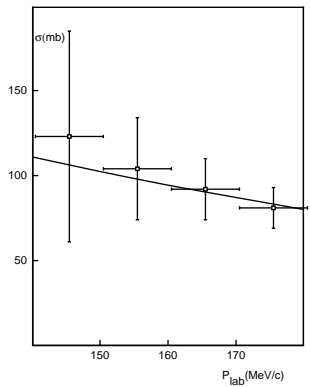}
\caption{The $\Sigma^{+}p$ elastic total cross section~\protect\cite{Ei71}.}
\label{fig.4}
\end{figure}

For elastic $\Sigma^{+}p$-scattering (see Figure~\ref{fig.4}) there exist
4 datapoints~\cite{Ei71} in the momentum range 
145 MeV/c $< p_{\rm L} < 175$ MeV/c which corresponds to \\
9 MeV $< T_{\rm lab} < 13$ MeV. 

\begin{figure}[bt]
\vspace{6cm}
\includegraphics{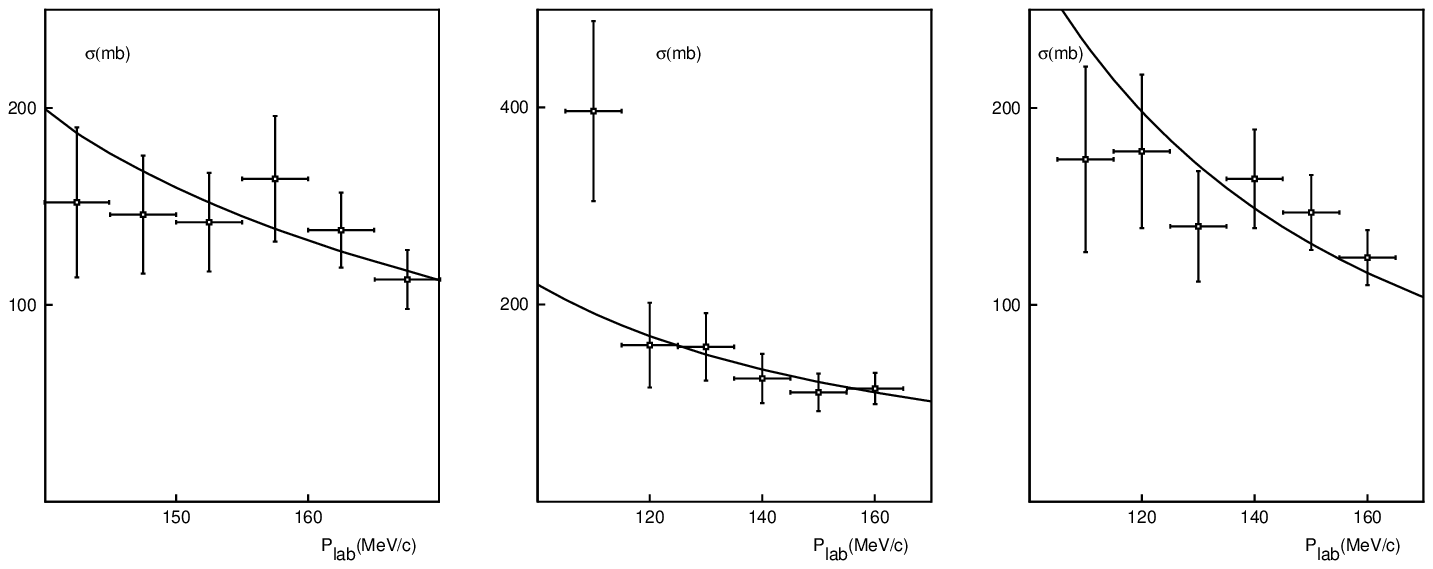}
\caption{The total elastic cross sections $\Sigma^{-}p
  \rightarrow \Sigma^{-}p$ and the charge exchange reactions $\Sigma^{-}p
  \rightarrow \Sigma^{0}n$ and $\Sigma^{-}p \rightarrow \Lambda^{0}n$ 
 \protect\cite{Ei71,En66}.}
\label{fig.5}
\end{figure}

For elastic $\Sigma^{-}p$-scattering, and the charge exchange reactions
$\Sigma^{-}p \rightarrow \Sigma^{0}n$ and $\Sigma^{-}p \rightarrow \Lambda n$ 
one has for each reaction 6 datapoints~\cite{Ei71,En66}
in the momentum interval 142 MeV/c $< p_{\rm L} < 168$ MeV/c
or 9 MeV $< T_{\rm L} < 12$ MeV (see Figure~\ref{fig.5}).  

The restricted
dataset contains finally also the ratio at rest $r_{R}$ from the production
of $\Sigma^{0}$ and $\Sigma^{0}$ and $\Lambda^{0}$ hyperons, when stopped
$\Sigma^{-}$ hyperons are captured by protons~\cite{He68}. 
This ratio $r_{R} = \Sigma^{0}/(\Sigma^{0}+\Lambda) = 0.468(10)$ 
is one of the few numbers in these reactions
with a rather good accuracy. \\[2mm]

It is clear from this dataset, that the data are really scarce and that
they have large errors. Because of the low energies these data contain mainly
$s$-wave information. The allowable number of parameters is of the order of 6,
one for each of the 5 reactions and one for the ratio at rest. 

More recently there have come available beautiful data~\cite{Ba87} for the
strangeness exchange reaction $\bar{p}p \rightarrow \bar{\Lambda}\Lambda$.
These data are taken at the laboratory momenta
$p_{L} < 1.55$ GeV/c, which corresponds to the energy
$E$ in the center of mass,
\[
E = \sqrt{s} - 2m_{\Lambda} \leq 39.1\ {\rm MeV}\ .
\]
Available at present are $N_{d} = 157$ datapoints corresponding to 99
differential cross sections, 38 polarizations, 20 spin-correlations.
This part of the database is rapidly growing. There are also already some
measurements available \cite{Ba90} of the reactions
\[
\bar{p}p \rightarrow \Lambda\bar{\Sigma},\ \Sigma\bar{\Lambda},\
 \Sigma\bar{\Sigma},\ {\rm etc.}
\]

\section{Flavor SU(2): Isospin}  \label{III}
Isopin symmetry is a good symmetry in the $Y\!N$-interactions, when 
the Coulomb interactions can be neglected and when there
are no important mass differences between particles of the same
isomultiplet. So when no $(n,\ p)$, and $(\Sigma^{+},\ \Sigma^{0},\ \Sigma^{-})$
mass differences are taken into account. The most important manifestation
of this approximation is the coincidence of the  various $\Sigma N$-%
thresholds. Also the Coulomb interaction in the $\Sigma^{+}p$ and
$\Sigma^{-}p$ channels should be neglected. 

However, it is important to take the breaking of the isospin symmetry
of SU(2) flavor into account. The Coulomb interaction in the $\Sigma^{-}p$-%
channel is very important for the ratio $r_{R}$ at rest, because the
attractive Coulomb interaction enhances the strong reaction rates. 
The Coulomb interaction
manifests itself also in the differential cross sections $d\sigma/d\Omega$
for the elastic scatterings $\Sigma^{-}p$ and $\Sigma^{+}p$. 

The mass differences between members of the same isomultiplet are another 
source of breaking of the isospin symmetry. A first manifestation of
this is the presence of the different $\Sigma N$ thresholds
\[
E_{\rm th} (\Sigma^{0}p) - E_{\rm th}(\Sigma^{+}n) = 1.8\ {\rm MeV} 
\rule{5mm}{0mm} {\rm and} \rule{5mm}{0mm}
E_{\rm th} (\Sigma^{-}p) - E_{\rm th}(\Sigma^{0}n) = 3.6\ {\rm MeV} 
\]
The mass difference between the pions $(\pi^{0},\pi^{\pm})$ gives rise to
different interaction strength and different ranges.

In the $\Lambda N$ interaction there is a very interesting kind of 
isospin-breaking~\cite{Da64,Sw71}. 
The isospin of the $\Lambda$ hyperon is $I=0$ and this
forbids one-pion-exchange in the elastic $\Lambda N$-scattering.
OPE gives rise to the reaction $\Lambda N \rightarrow \Sigma N$.
However, the physical $\Lambda$ is {\bf not} a pure $I=0$ state. Due to the
electromagnetic interaction the $\Lambda$ has a small $\Sigma^{0}$
component mixed in, such that
\[
\Lambda_{\rm phys} = \cos\theta\ \Lambda + \sin\theta\ \Sigma^{0}
\rule{5mm}{0mm} {\rm and} \rule{5mm}{0mm}
\Sigma^{0}_{\rm phys} = - \sin\theta\ \Lambda + \cos\theta\ \Sigma^{0}
\]
The pion $\pi^{0}$, which does not couple to the bare $\Lambda$,
couples to the physical $\Lambda_{\rm phys}$, because it couples to 
$\Sigma^{0}$. 
Because the coupling constants of the 
$\pi^{0}$ to the proton and to the neutron have opposite sign,
there is an isospin symmetry breaking due to this one-pion-exchange.
The result is a rather weak, but noticeable, isospin breaking,
OPE-potential in the $\Lambda N$ channel. 

\section{Flavor SU(3)}  \label{IV}
An important manifestation of SU(3) flavor symmetry~\cite{Sw63}
and the quark model~\cite{Ko69} is the appearance of
mesons in {\bf nonets}. Important nonets are the $J^{PC}=0^{-+}$
pseudoscalar meson nonet, the $J^{PC}=1^{--}$ vector meson nonet, and the
$J^{PC}=0^{++}$ scalar meson nonet. The
non-strange members of these nonets are $(\pi,\eta,\eta^{\prime})$,
$(\rho,\omega,\phi)$, and $(a_{0}(980),\ f_{0}(975),\ f_{0}(760))$. The strange 
members in each nonet appear in two isodoublets with $Y=\pm 1$, $(K^{+},
K^{0})$ and $(\overline{K^{0}}, K^{-})$. They are the pseudoscalar
$K(495)$, the vector $K^{*}(892)$, and the scalar $\kappa(880)$.

The baryons appear mainly in octets $\{ \ul{8}\}$, decuplets
$\{ \ul{10}\}$ and singlets $\{ \ul{1}\}$. The most important example
being the $J^{P}=\frac{1}{2}^{+}$-baryon octet.

When one wants to place the deuteron in an SU(3) multiplet, then this 
\cite{Oa63}
must be an anti-decuplet $\{ \ul{10}^{*}\}$. This multiplet contains
presumably also the state with $Y=1$, $I=\frac{1}{2}$ and mass $M=2129$ MeV
near the $\Sigma N$-threshold. The equal spacing rule predicts
then a $\Xi N$, $\Lambda\Sigma$ and $\Sigma\Sigma$ resonance with
$Y=0$, $I=1$ and mass $M=2382$ MeV near the $\Sigma\Sigma$ threshold and a
state with $Y=-1$, $I=\frac{3}{2}$ around $M=2635$ MeV.

The breaking of the SU(3) symmetry in the baryon masses has a 
noticeable effect. 

\vspace{\baselineskip}

For the description of the $BB$-interaction in general the SU(3) flavor 
symmetry is useful as a limiting case~\cite{Sw71,Iw64,So64,Do90}. 
The $J^{P}=\frac{1}{2}^{+}$ 
baryons $(N,\Sigma,\Lambda,\Xi)$ all belong to the $\frac{1}{2}^{+}$-baryon
octet. The flavor wave function of the two-baryon states must belong
to one of the SU(3) irreps contained in the right-hand-side
of the SU(3) Clebsch-Gordan series:
\[
\{ 8\} \times \{ 8\}  = \underbrace{\{ 27\}  + \{ 8\} _{S} + \{ 
        1\} }_{\rm symmetric} +
     \underbrace{\{ 10\}  + \{ 10^{*}\}  + \{ 8\} _{A}}_{\rm anti-symmetric}\ .
\]
The symmetry of the flavor wave function under interchange of the two
baryons is indicated. The total wave function $\psi$ can be written
as the product of a space-, a spin-, and a flavor-wave-function:
\[
\psi = ({\rm space})({\rm spin})({\rm flavor})\ .
\]
The generalized Pauli principle requires that the total wave function
$\psi$ is antisymmetric under interchange of the two baryons. This implies
for the flavor symmetric states $\{ 27\}$, $\{ 8\}_{S}$ and $\{ \ul{1}\}$
the antisymmetric space-spin combinations $^{1}S_{0}$, $^{3}P_{1}$, $^{1}D_{2}$,
$^{3}F$, etc, and for the flavor anti-symmetric states $\{ \ul{10}\}$, 
$\{ \ul{10}^{*}\}$ and $\{ 8\}_{A}$
the symmetric space-spin combinations $^{3}S_{1}$, 
$^{1}P_{1}$, $^{3}D$, $^{1}F_{3}$, $^{3}G$, etc.  \\

%
%

\hspace*{-0.8cm}\begin{tabular}{lll}
\multicolumn{3}{l}{\rule{1cm}{0mm} The $N\!N$-states have $Y=2$ and 
  $I=0$ and 1:} \\[0.5mm]
the $I=1$ states $^{1}S_{0}$, $^{3}P$, $^{1}D_{2}$, $^{3}F$, etc. & 
  belong to & $F=\{ 27\}$,  \\[0.5mm]
the $I=0$ states $^{3}S_{1}$, $^{1}P_{1}$, $^{3}D$, $^{1}F_{3}$, etc. & 
  belong to & $F=\{ \ul{10}^{*}\}$.  \\[2mm]
\multicolumn{3}{l}{\rule{1cm}{0mm} The $YN$-states have $Y=1$ and 
  $I=\frac{1}{2}$ and $\frac{3}{2}$:} \\[0.5mm]
the $I=\frac{3}{2}$ states $^{1}S_{0}$, $^{3}P$, $^{1}D_{2}$, $^{3}F$,
  etc.  & belong to & $F=\{ 27\}$,  \\[0.5mm]
the $I=\frac{3}{2}$ states $^{3}S_{1}$, $^{1}P_{1}$, $^{3}D$, $^{1}F_{3}$,
  etc.  & belong to & $F=\{ \ul{10}\}$.  \\[2mm]
\multicolumn{3}{l}{\rule{1cm}{0mm} The situation for the 
  $I=\frac{1}{2}$ states is more complicated:} \\[0.5mm]
the $I=\frac{1}{2}$ states $^{1}S_{0}$, $^{3}P$, etc.\ belong to &  
  a mixture of & $F=\{ 8\}$ and $\{ 27\}$,  \\[0.5mm]
the $I=\frac{1}{2}$ states $^{3}S_{1}$, $^{1}P_{1}$, etc. belong to & 
  a mixture of & $F=\{ 8\}$ and $\{ 10^{*}\}$. 
\end{tabular}

\section{$N\!N$-models}  \label{V}
Let us give a quick review of some of the $N\!N$-models that appear in
the literature. In Nijmegen we have constructed various $N\!N$-potential
models. They are
\begin{itemize}
\item {\bf hard core} models A to F. The models A and B stem~\cite{Na73a,Na73b}
  from 1973, the model D~\cite{Na75,Na77} from 1975--1977, 
  and the model F~\cite{Na79} from 1979.
\item {\bf soft core} models. The Nijmegen soft-core model (Nijm78)
based on Regge-trajectory exchange~\cite{Na78}
stems from 1978. The corresponding
$Y\!N$-model~\cite{Ma89} was constructed in 1989. Recently the 
$N\!N$-model~\cite{Na78} has been updated~\cite{potmod93}. 
This updated version Nijm93 has $\chi^{2}$/datapoint = 1.87
with respect to all the available $N\!N$-scattering data below $T_{L}=350$ MeV.
\item {\bf extended soft core} (ESC) model. This 1993 
soft core model~\cite{Rij93}, \linebreak[4] inspired by chiral-symmetry,
gives a fit to the available $N\!N$-data with \linebreak[4]
$\chi^{2}$/datapoint = 1.16 (17 MeV $\leq T_{\rm lab} \leq 350$ MeV). 
The corresponding $Y\!N$-potentials have not been constructed yet.
\item {\bf Reidlike} models. In 1993 several Reidlike models, NijmI and
 NijmII, based on the Nijm78 potential, have been constructed~\cite{potmod93}. 
Also an update Reid93 of the old Reid soft-core 
potential (RSC) was constructed~\cite{potmod93}.
These potentials have all excellent fits with respect to 
the $N\!N$-data; they all have \linebreak[4] $\chi^{2}$/datapoint = 1.03.
\end{itemize}

The Paris $N\!N$-potential~\cite{La80} Paris80 has a fit with the $N\!N$-data
that is comparable with the old Nijm78 potential.

Also in Bonn one has constructed various $N\!N$-potentials. This started
with potentials like HM1976~\cite{Bo77}. In 1987 various Bonn $N\!N$-potentials
were published~\cite{Bo87} with various names. 
In 1989 the Bonn group constructed a special $pp$-potential~\cite{BoPRC}
and again several $N\!N$-potentials, like BonnA,
B, C~\cite{BoPRep}, and from each one several versions.

\section{Models for the $Y\!N$- and $Y\!Y$-interaction}  \label{VI}
Of the various models that exist for describing the $Y\!N$- and the
$Y\!Y$-interaction, there are first of all the hard core models Nijmegen D
and F. The main difference between these two models is the treatment of
the scalar mesons. In D an SU(3) singlet is assumed and in F an SU(3) nonet.
This model F was extended~\cite{Mac71} to the $Y=0$ channels $\Lambda\Lambda$,
$\Xi N$, etc.

The Nijmegen soft-core model Nijm78 for the $N\!N$-interaction was
extended~\cite{Ma89} to the $Y\!N$-interaction in Nijm89.

Another generalization exists, called the SCW-model~\cite{Mae}. In this model,
to the meson theoretical interaction is added in every SU(3) channel
either a repulsive soft-core or an attractive soft well.  The
resulting potential has been generalized to the $Y=0$ channels by
P. Maessen et al.
With this model an excellent fit to the $Y\!N$-data has been obtained
with only a few parameters (see figures~\ref{fig.3} to \ref{fig.5}). 

Also a boundary condition model was constructed~\cite{Mae}. 
Here the meson-theoretic
potential was used to describe the interaction for values of $r > 1.4$ fm.
At the radius $r=1.4$ fm was specified the boundary condition
\[
P = b (d\psi/dx)/\psi\ .
\]

\section{OBE-part of the meson-theoretical potentials} \label{VII}
It has already been stated that the mesons come in {\bf nonets}
$[9]$, where \linebreak[4]
$[9] = \{ \ul{8}\} \oplus \{ \ul{1}\}$ and $\{ \ul{8}\}$ and 
$\{ \ul{1}\}$ are an SU(3) octet and singlet. In the $N\!N$-channels
are exchanged from each nonet: \\
(i) one meson with $Y=0$, $I=1$ like $\pi, \rho, a_{0}$, and \\
(ii) two mesons with $Y=0$, $I=0$ like $\eta, \eta^{\prime}, \omega, \phi$, and
$f_{0}(760)$, $f_{0}(975)$. \\
When one wants to describe also the $Y\!N$-channels,
then one needs to consider also the exchange of the $Y=\pm 1$ $I=\frac{1}{2}$
strange mesons like $(K^{+} K^{0}), (\overline{K^{0}}, K^{-})$.

The $I=0$ mesons are mixed due to for example
the SU(3) breaking of the quark masses. 
The mixing angle $\theta$ is introduced to describe this mixing.
For the pseudoscalar mesons one writes
\begin{eqnarray*}
\begin{array}{lcrcl}
\eta & = & \eta_{8} \cos\theta_{ps} & - & \eta_{1} \sin\theta_{ps} \\
\eta^{\prime} & = & \eta_{8} \sin\theta_{ps} & + & \eta_{1} \cos\theta_{ps} 
\end{array}
\end{eqnarray*}
{}From the linear Gell-Mann-Okubo mass formula one predicts 
$\theta_{ps} = -23$ degrees. Using the quadratic mass formula one
gets $\theta_{ps} = -10.1$. Experimentally is seems to be that
$\theta_{ps} \sim -20$ degrees. This does not imply that the linear
GMO mass formula is better, because the
mixing angle is very sensitive to small corrections.

\section{Pseudoscalar mesons}  \label{VIII}
The coupling of the pseudoscalar mesons $J^{PC}=0^{-+}$ with the $J^{P}=
\frac{1}{2}^{+}$ baryons can be described by \\
either \rule{5mm}{0mm} PS-coupling:\ \  ${\cal L}_{PS} = 
  g (\overline{\psi}\gamma_{5}\psi)\phi$ \\
or \rule{1cm}{0mm} PV-coupling:\ \  ${\cal L}_{PV} = \frac{f}{m_{s}} 
       ( \overline{\psi} \gamma_{\mu} \gamma_{5} \psi) \partial^{\mu}\phi$.

In the PV-lagrangian a scaling mass $m_{s}$ is introduced in order to make the 
coupling constant $f$ dimensionless. We feel that one must always choose
the same mass e.g.\ $m_{s} = m(\pi^{+})=m_{+}$ for this scaling mass.
The coupling constants we will denote by $f$ in that case. 

Some people prefer to take the scaling mass equal to the mass of
the exchanged meson $m_{s}=m_{\phi}$. The coupling constants we will
denote in that case by $f^{\prime}$.

For the pseudo-scalar-meson-baryon-baryon vertex 
there exist an equivalence between PS- and PV-coupling constants:
\[
f^{2} = \left( \frac{m_{s}}{M_{1}+M_{2}} \right)^{2} g^{2}\ .
\]
The coupling constant $f_{p}$ of the $\pi^{0}$ with the proton~\cite{St93} is
$f_{p}^{2}=0.075$. This corresponds to $g^{2}=13.56$. This same
$g^{2}=13.56$  corresponds for this $\pi^{0}pp$ vertex to ${f^{\prime}}^{2}
=0.070$. The question arises now: ``Which of these coupling constants
$f$, $f^{\prime}$, or $g$ is approximately SU(3) symmetric?''
It is clear that when the PV-coupling constants $f$ are approximately
SU(3) symmetric, that then the PS-coupling constants $g$ and the
PV-coupling constants $f^{\prime}$ have a sizeable SU(3) breaking. 

When one assumes SU(3) for the PV-coupling constants $f$ then the Cabibbo
theory of the weak interactions and the Goldberger-Treiman relation 
predict the value~\cite{Du83} $\alpha_{PV} = (F/(F+D))_{PV}=0.355(6)$.
This value was also found in~\cite{Ma89} while fitting the $Y\!N$-data.
In a study~\cite{Ti91} of the reaction $\bar{p}p\rightarrow 
\bar{\Lambda}\Lambda$ Timmermans et al.\ found either $\alpha_{PV}=0.34(4)$ or
$\alpha_{PS} = 0.42(4)$. The agreement between the two values of $\alpha_{PV}$
indicates a preference for PV-coupling.

For a complete description of the coupling of the PS-mesons to the baryons
we need to know the mixing angle $\theta$, the singlet coupling constant
$f_{1}$, the octet coupling constant $f_{8}$ and the ratio
$\alpha_{PS}=F/(F+D)$. However, this is not all. There is still the
question: What is better, SU(3) symmetry for the PS or for the PV coupling
constants?

These coupling constants are just phenomenological parameters. The
spatial extension  of the baryons and the mesons introduces a form 
factor~\cite{Sw77}.
In first approximation it is assumed that the coupling constants become
dependent on the momentum transfer. Then the question arises, where do
we assume SU(3) for the coupling constants? At the pole or at $t=0$?
When the values at the meson pole are assumed to be SU(3) symmetric,
then the values of the coupling constants at $t=0$ will in general
{\bf not} be SU(3) symmetric anymore and vice versa.

As far as the specific value of the coupling constant is concerned it is
interesting to note in 1993 that the value of the $\pi N\!N$ coupling
constant deduced for the model D in 1975 was $g^{2}=13.4$ or 
$f^{2}=0.074$~\cite{Na75}.

\section{The Vector mesons}  \label{IX}
The $J^{PC} = 1^{--}$ vector meson nonet contains the non-strange mesons
$\rho,\omega$, and $\phi$. The $Q\overline{Q}$-quark model SU(3) eigenstates
are
\begin{eqnarray*}
\omega_{8} & = & [ u\bar{u} + d\bar{d} -2 s\bar{s}]/\sqrt{6} \\
\omega_{1} & = & [ u\bar{u} + d\bar{d} + s\bar{s}]/\sqrt{3} 
\end{eqnarray*}
When these states are ideally mixed, then
\begin{eqnarray*}
\begin{array}{lcrclcl}
\omega & = & \cos \theta_{v}\ \omega_{1} & + & \sin\theta_{v}\ \omega_{8} & =
   & [ u\bar{u} + d\bar{d}]/\sqrt{2} \\
\phi & = & -\sin \theta_{v}\ \omega_{1} & + & \cos\theta_{v}\ \omega_{8} & =
   & -s\bar{s}
\end{array}
\end{eqnarray*}
This ideal mixing angle has then
\[
\sin\theta_{v} = 1/\sqrt{3}\ , \ \ \tan\theta_{v} = 1/\sqrt{2}\ ,\ \ 
  {\rm and}\ \ \theta_{v} = 35.26\ .
\]
The physical coupling constants are related to the coupling constants 
$g_{\omega_{8}}$ and $g_{1}$ of the unmixed states. Then
\begin{eqnarray*}
\begin{array}{lcrcl}
g_{\phi} & = & - \sin\theta_{v}\ g_{1} & + & \cos\theta_{v}\ g_{\omega_{8}} \\
g_{\omega} & = & \cos\theta_{v}\ g_{1} & + & \sin\theta_{v}\ g_{\omega_{8}} 
\end{array}
\end{eqnarray*}
The OZI-rule~\cite{OZI} states that the $\phi$-meson is in first approximation not coupled
to the nucleons. Thus $g_{\phi} =0$. This implies then that
\[
g_{1} = \sqrt{2} g_{\omega_{8}} \ \ {\rm and}\ \ g_{\omega} = \sqrt{3} 
  g_{\omega_{8}}\ .
\]
The coupling constants $g_{\omega_{8}}$ is related to the $\rho$-coupling
constant $g_{\rho}$ by
\[
g_{\omega_{8}} = [(4\alpha_{v}{-1})/\sqrt{3}] g_{\rho} = \sqrt{3} g_{\rho}\ .
\]
In the last step above we used Sakurai's idea that the vector mesons are
universally coupled~\cite{Sa60}. 
This requires $\alpha_{v}=1$. The coupling constant
$g_{\omega}$ is therefore related to the $\rho$-coupling constant 
$g_{\rho}$ by $g_{\omega}^{2} = 9 g_{\rho}^{2}$.

In treating the vector mesons there are still many uncertainties. The
couplings to the $J^{P} = \frac{1}{2}^{+}$ baryons are described either
by the Dirac and the Pauli coupling constants or by the electric and the
magnetic coupling constants. For which of these holds SU(3)?
Again one comes up with the question whether one needs the coupling
constants at the particle poles or at $t=0$.

The ratio $(f/g)_{\rho}$ is rather controversial. Vector meson dominance (VMD)
predicts~\cite{Sa69} $(f/g)_{\rho} = 3.7$. From analyses of the $\pi N$ data the
Karlsruhe people~\cite{Ho75} determined 
long ago that $(f/g)_{\rho}=6.1$, but also that $g^{2}=14.28$ for the
$\pi N\!N$-coupling constant. It appears
that the $\pi N$ data available at the time of the Karlsruhe analyses were
not so great. In Nijmegen we made a fit to the $N\! N$-scattering data to
determine $(f/g)_{\rho}$. In Nijm78 we found $(f/g)_{\rho}=4.3$. Recently
this potential was refitted and now we find that $(f/g)_{\rho}=4.1$.
We see that the Nijmegen determination is close to the VMD value. \\
Popular values for the $F/(F+D)$-ratio's $\alpha_{E}$ and $\alpha_{M}$ are\\
\rule{1cm}{0mm}$\alpha_{E}=1$ from universal coupling \`a la 
Sakurai~\cite{Sa60}, and \\
\rule{1cm}{0mm}$\alpha_{M}=0.275$ using relativistic SU(6) 
(Sakita and Wali~\cite{Sak65}).

\section{The scalar mesons}  \label{X}
The scalar meson $\sigma$, the fictitious $\sigma$, with a mass of about
$M \sim 550$ MeV was introduced in 1960--1962 by N. Hoshizaki 
et al.~\cite{Ho60} and
used in 1964 by Bryan and Scott~\cite{Br64}. 
This scalar meson was required in OBE-models
for $N\!N$ in order to get \\
(i) intermediate range attraction, and \\
(ii) sufficiently strong ${\bf L}\cdot{\bf S}$-forces.

In $\pi\pi$ production experiments there often appeared a broad structure
$\varepsilon(760)$ under the $\rho^{0}$. Because the signal of the 
$\rho^{0}$ is so strong, the existence of the broad structure was always
unsure. 

The $\pi\pi$ interaction is traditionally studied in the reaction
$\pi N\rightarrow \pi\pi N$. When the production of the pion goes
via pion exchange, we have a $\pi\pi \rightarrow \pi\pi$ vertex and here
$\pi\pi$ scattering has been studied. In this scattering sometimes
$\varepsilon(760)$ appeared as an established particle, sometimes its
existence was denied. In a recent analysis~\cite{Sv92} of this 
production reaction
also the exchange of other mesons, besides the pion, was assumed. In this 
recent analysis the $\varepsilon$-meson has mass $M=750$ MeV and width
$\Gamma \sim 100-150$ MeV.

An important development in the treatment of the scalar mesons was the
realization~\cite{Bi71} in 1971 that 
the exchange of a wide $\varepsilon(760)$
simulates the exchange~\cite{Sw77} of the fictitious, 
low mass $\sigma$. The potential
due to the exchange of a wide $\varepsilon$ can be calculated~\cite{Sch71},
where the $2\pi$ threshold is taken properly into account. For easy handling, 
necessary in the older computers, the potential $V(\varepsilon)$ of the wide
epsilon was approximated as the sum of two Yukawa's. One of these
Yukawa's has a low mass and the other one a high mass. In the Nijm93 
potential these masses are $m_{\rm low} = 488$ MeV and $m_{\rm high} = 1021$
MeV.

The $Q\overline{Q}$-mesons with $J^{PC}=0^{++}$ must belong to the 
$^{3}P_{0}$-states. The assignments of the $2^{++}$ and  the $1^{++}$
mesons are generally accepted. The assignments of the scalar mesons
are more controversial. However, let us start with the masses 
$2^{++}$ and $1^{++}$ mesons.
\[
\begin{array}{lllll}
^{3}P_{2}, & J^{PC}=2^{++}; & a_{2}(1320), & f_{2}(1270), & f^{\prime}_{2}
  (1525) \\
^{3}P_{1}, & J^{PC}=1^{++}; & a_{1}(1260), & f_{1}(1285), & f^{\prime}_{1}
  (1510) \ ,
\end{array}
\]
and predict the masses of the $0^{++}$-mesons
\[
\begin{array}{lllll}
^{3}P_{0}, & J^{PC}=0^{++}; & a_{0}(1300), & f_{0}(1300), & f^{\prime}_{0}
  (1500) \ .
\end{array}
\]
The Particle Data Group~\cite{PDG92}
lists an $a_{0}(1320)$, which needs confirmation and various $f_{0}$'s. The 
predicted masses look reasonable. What about the scalar-mesons
\[
\delta(980),\ S(975),\ \ {\rm and}\ \ \varepsilon(760)\ ?
\]
One notices first of all the non-familiar mass relation
\[
m(a_{0}) \approx m(f_{0}) \gg m(f_{0}^{\prime})\ .
\]
This mass relation is just contrary to the mass relation of the $^{3}P_{J}-
Q\overline{Q}$-mesons, where
\[
m(a_{J}) \simeq m(f_{J}) \ll m(f_{J}^{\prime})\ .
\]
The non-familiar mass relation (low $f^{\prime}$ mass)
 is easily understood from the quark content. The
$Q\overline{Q}$-mesons $a_{J}$ and $f_{J}$ contain only non-strange 
quarks and the heavier $Q\overline{Q}$-meson $f_{J}^{\prime}$ contains the
strange quarks $(s\bar{s})$.

In a $Q\overline{Q}$-model in 1980 Aerts et al.~\cite{Ae80} predicted for the
mesons with the non-strange quark content $n\bar{n}$ a mass around
1285 MeV. The $I=0$ and $I=1$ mesons being almost degenerate. The $I=0$ meson
with $s\bar{s}$ content was predicted around $M=1475$ MeV.

A solution for this non-familiar problem in the quark model was given in 1977 by
R.L. Jaffe~\cite{Ja77}. 
He calculated in the MIT-bagmodel the $q^{2}\bar{q}^{2}$ states.
The lowest states were a {\bf nonet of scalar mesons}.
A heuristic treatment runs as follows. The lowest $q^{2}$ states is a diquark 
with $F=3^{*}$, $C=3^{*}$, and $S=0$, where $F$ is flavor, $C$ is color
and $S$ is spin. Because of the $F=3^{*}$ assignment we will denote
these states by $\overline{Q}$. Thus
\[
\overline{Q} = \begin{array}{ccc}
    & \overline{S} & \\[5mm]
    \overline{U} & & \overline{D} 
    \end{array}  =
 \begin{array}{ccc}
    & [ud] & \\[5mm]
    \ [sd] & & [su] 
    \end{array} 
\]
With $[ud]$ we mean the antisymmetric flavor wave function $ud-du$.
The lowest $q^{2}\bar{q}^{2}$ states are formed from $\overline{Q}$,
an antitriplet, and the antiparticles $Q$, a flavor triplet. The 
$\overline{Q}Q$ combination is a flavor nonet. 
The lowest mass state 
\[
\overline{S}S = [ud][\bar{u}\bar{d}]
\]
is an $I=0$ scalar meson containing only non-strange
quarks with predicted~\cite{Ja77}
mass $M=690$~MeV. This is the $f_{0}^{\prime}$ meson of this nonet.

There exist in this nonet also a degenerate pair of $I=0$ and $I=1$ mesons.
The neutral mesons (think of the $\rho^{0}$ and $\omega^{0}$) are
\[
(U\overline{U} \pm D\overline{D})/\sqrt{2} = \{ [\bar{s}\bar{d}]
  [sd] \pm [\bar{s}\bar{u}][su]\}/\sqrt{2}\ .
\]
The predicted mass was $M=1150$ MeV. It is obvious that these mesons are the
$f_{0}(975)$ and $a_{0}(980)$ at the $K\overline{K}$-threshold. From the
wave function we see that these mesons contain an $\bar{s}s$-pair.
This explains for example why the $f_{0}(975)$ meson with a mass
{\bf below} the $K\overline{K}$-threshold can decay for about 22\% into a
$K\overline{K}$-pair.

The strange partner, called $\kappa$, must have flavor wave functions like
\[
[ud][\bar{s}\bar{d}] \ \ {\rm and}\ \ [ud][\bar{s}\bar{u}]\ , \ \ {\rm etc}
\]
These mesons contain only one $s$ or one $\bar{s}$. The expected mass
is around 880~MeV, just under the strong signal of the
$K^{*}(892)$. This explains why the scalar meson $\kappa$ is so hard to detect.
This meson has been seen by Svec~\cite{Svec92} in 1992 with mass $M=887$ MeV.

How to describe the mixing of these scalar mesons? We write
\begin{eqnarray*}
\begin{array}{lcrcl}
f_{0}^{\prime} & = & \cos\theta_{s}\ \varepsilon_{1} & + & \sin\theta_{s}\ 
  \varepsilon_{8} \\
f_{0} & = & -\sin\theta_{s}\ \varepsilon_{1} & + & \cos\theta_{s}\ 
  \varepsilon_{8} 
\end{array}
\end{eqnarray*}
When we assume ideal mixing, then
\[
f_{0}^{\prime} = \varepsilon = S\overline{S} \ \ {\rm and}\ \ 
 f_{0} = S = (U\overline{U}+D\overline{D})/\sqrt{2}\ ,
\]
which means $\tan\theta_{s} = -\sqrt{2}$ and $\theta_{s}=\theta_{v}-90=
-54.75$. However, this is not the only mixing present. One expects also mixing
with the $q\bar{q}$ $(^{3}P_{0})$ states, with glueballs, etc.

\section{The Pomeron}  \label{XI}
In the region above $p_{\rm lab} = 2$ GeV/c boson exchange has to be replaced
by Reggeon exchange, because the total cross section becomes there 
approximately constant, see e.g.~\cite{Sw88}. This feature of the total
cross section can only be explained in the Regge pole model. It was
pointed out in~\cite{Sw88} that in Regge pole models, see for 
example~\cite{Ba66}, the pomeron gives a very significant contribution
already at $p_{\rm lab} = 2$ GeV/c. At this momentum $\sigma_{T} \approx
45$ mb and $\sigma_{\rm el} \approx 20$ mb. When the pomeron is omitted,
the model of~\cite{Ba66} would predict $\sigma_{\rm el} \approx 2.3$ mb.

In low energy pion-nucleon and kaon-nucleon scattering the presence of the
pomeron has been demonstrated using finite-energy sum rules~\cite{Do68}.
There it appeared that, after the subtraction of the baryon resonances,
the remaining background amplitude is directly related to pomeron-exchange.
This background amplitude is important for the scattering lengths.

Since the Regge-region, $p_{\rm lab} > 2$ GeV/c, is not that remote
from the $N\Delta$-region, $p_{\rm lab} \approx 1.32$ GeV/c, or even the
low-energy region, $p_{\rm lab} < 0.9$ GeV/c, it is essentially the same
physics that governs the low-energy and the Regge-regions. A unified
description of these regions is therefore desirable.

A unification for the baryon-baryon channels, using the
Khuri-Jones representation~\cite{Jo62} of the Regge poles, has been
worked out~\cite{Rij75} and applied by the Nijmegen group to baryon-baryon
scattering~\cite{Na78,Rij75}.
Phenomenologically, the inclusion of the pomeron-exchange potential
in these models serves to give reasonable values for the $\omega$-coupling.
More fundamentally, in these OBE-models, where there are, of course,
no $N\overline{N}$-pair contributions to the potentials, the strong $\varepsilon
N\!N$-coupling would be at variance with the small $s$-wave pion-nucleon
scattering lengths. The pomeron contribution helps out here by
cancelling largely this $\varepsilon$-contribution.

The physical picture of the pomeron has changed over the years in accordance
with the progress of our understanding of the hadrons. In the sixties
and early seventies, the pomeron was associated with multiperipheral 
chains. This is natural in chiral theories, where one envisions a cloud of soft
pions around constituent quarks. With the advent of QCD, ons tries to
explain most of the pomeron features by considering it as a two-gluon
(or multigluon) system~\cite{Lo75}.

The repulsive character of the pomeron-potential appears often a little
puzzling, because at low energy it is very similar to scalar-exchange
and one would therefore expect an attractive potential. The repulsiveness
comes from the Regge phenomenology. The pomeron residue is positive,
because pomeron-exchange is directly related to $\sigma_{T}$ at high-energy.
In the Khuri-Jones procedure this then leads directly to a repulsive 
potential. To look for a more detailed explanation we examine the
two-gluon picture. First of all, we assume that the pomeron couples
primarily to the quarks, as indicated by high-energy experiments~\cite{He92}.
This has been related to the QCD-vacuum properties~\cite{La87}. The pomeron
quark-coupling picture restores the ``additive quark rules'' for the
pomeron~\cite{Ko69}. This in contrast to the so-called ``subtracted''
quark-picture, where in the two-gluon coupling to a hadron one sums
independently over the quark couplings of the individual gluons~\cite{Pu81}.
In this latter picture the pomeron-exchange potential would be due
to the effects of induced color-electric dipoles, like van der Waals forces.
These would then most likely be attractive.

Assuming that the Coulomb part of the two gluons in pomeron-exchange dominates
the interaction, it is not unrealistic to consider for the pomeron
quark-quark potential a two-scalar exchange model. It is well-known~\cite{Sa51}
that then in the adiabatic approximation all contributions cancel. 
The first non-vanishing contribution to the potential comes from
non-adiabatic corrections. This gives rise to a repulsive potential 
between the quarks of the form
\[
V_{P_{qq}}(r) \sim g^{2}_{P_{qq}} (\Lambda r)^{2} \exp \left[ - \frac{1}{2}
  (\Lambda r)^{2} \right]\ .
\]
The same $V_{P_{qq}}(r)$ can be derived in the context of QCD, relating the 
strength of the potential to the vacuum expectation $\langle 0|G_{\mu\nu a}
(x) G^{\mu\nu a}(y)|0\rangle$~\cite{Rij94}.
Folding the $V_{P_{qq}}$ potential with the baryon quark-model wave functions, 
one arrives at a repulsive pomeron-exchange $B\!B$-potential. Using gaussian
quark wave-functions gives a gaussian pomeron-exchange potential as used in
the Nijmegen models.

\section{The inner region}  \label{XII}
The treatment of the short range part of the interaction is very 
phenomenological. In the older models, like NijmD or NijmF we used hard cores.
In the $N\!N$-model Nijm78 and the corresponding $Y\!N$-model Nijm89
we used soft cores. 

Soft cores are generally introduced in the meson theoretic potentials, when
one uses form factors $F(k^{2})$, which cut down the high momentum
components sufficiently, such that the singularities at $r=0$ are
removed. In the Nijmegen soft core model we use exponential form factors
\[
F(k^{2}) = e^{-(k^{2}+m^{2})/\Lambda_{0}^{2}}\ .
\]
In the literature one uses mostly multipole form factors
\[
F(k^{2}) = \{ (\Lambda_{n}^{2} -m^{2})/(\Lambda_{n}^{2}+k^{2})\}^{n}
  \rule{1cm}{0mm} {\rm with} \ n=1,2,3,\ldots
\]
The advantage of the exponential form factor is that the coordinate space
potentials, obtained when using this form factor, are much softer
than when using the multipole form factors. 

Short-ranged are also the velocity dependent potentials of the form
\[
- \frac{\hbar^{2}}{2m} \{ \nabla^{2} \phi(r) + \phi(r) \nabla^{2} \}\ .
\]
Such potentials can be viewed as having introduced an $r$-dependent 
effective mass
\[
m_{\rm eff} = \frac{m}{1+2\phi(r)}\ .
\]

\vspace{2\baselineskip}

\noindent {\bf Acknowledgments} \\
We would like to thank both Dr.\ B. Gibson and Dr.\ R. Timmermans for
various discussions about these topics in the course of time.
Also the lively discussions with the members of our Nijmegen group and
with several of the participants of this pleasant US/Japan Seminar are
gratefully acknowledged.


\newpage
\begin{thebibliography}{99}
\bibitem{Sw71} J.J. de Swart, M.M. Nagels, Th.A. Rijken, and P.A. Verhoeven,
  {\it Springer Tracts Modern Physics} {\bf 60}, 138 (1971)
\bibitem{Sw88} J.J. de Swart, Th.A. Rijken, P.M. Maessen, and R.G. Timmermans,
  {\it Nuovo Cimento} {\bf 102 A}, 203 (1988) 
\bibitem{div} 
  J.J. de Swart, {\it Nukleonika} {\bf 25}, 397 (1980)\\
  J.J. de Swart and Th.A. Rijken, {\it Proceedings of the International
  Conference on Hypernuclear and Kaon Physics}, Heidelberg, 
  Germany (B. Povh, ed, 1982), pp.\ 271 \\
  J.J. de Swart and Th.A. Rijken, {\it Proceedings of the 1986 INS International
  Symposium on Hypernuclear Physics}, Tokyo, Japan (H. Band\={o}, et al., eds,
 1986), pp.\ 303 \\
 Th.A. Rijken, P.M.M. Maessen, and J.J. de Swart, {\it Nucl. Phys.} {\bf A 547},
  245c (1992) 
\bibitem{Na75} M.M. Nagels, Th.A. Rijken, and J.J. de Swart, {\it Phys. Rev.}
   {\bf D 12}, 744 (1975)
\bibitem{Na78} M.M. Nagels, Th.A. Rijken, and J.J. de Swart, {\it Phys. Rev.} 
   {\bf D 17}, 768 (1978)
\bibitem{Na79} M.M. Nagels, Th.A. Rijken, and J.J. de Swart, {\it Phys. Rev.} 
   {\bf D 20}, 1633 (1979)
\bibitem{Na77} M.M. Nagels, Th.A. Rijken, and J.J. de Swart, {\it Phys. Rev.} 
   {\bf D 15}, 2547 (1977)
\bibitem{Ma89} P.M.M. Maessen, Th.A. Rijken, and J.J. de Swart, {\it Phys. Rev.}
   {\bf C 40}, 2226 (1989)
\bibitem{La80} M. Lacombe, B. Loiseau, J.-M. Richard, R. Vinh Mau,
  J. C\^ot\'e, P. Pir\`es, and R. de Tourreil, {\it Phys. Rev.} {\bf C 21}, 861
  (1980)
\bibitem{Rij93} Th.A. Rijken, {\it Baryon-Baryon Interactions},
  {\it Proceedings of the XIVth European Conference 
  on Few-Body Problems in Physics}, Amsterdam, The Netherlands
  (B.L.G. Bakker, R. van Dantzig, eds, 1993)
\bibitem{Nijm93} V.G.J. Stoks, R.A.M. Klomp, M.C.M. Rentmeester, and J.J.
  de Swart, {\it Phys. Rev.} {\bf C 48}, 792 (1993)
\bibitem{Be86} A. Berdoz, F. Foroughi, and C. Nussbaum, {\it J. Phys. G} 
  {\bf 12}, L133 (1986)
\bibitem{Bo78} B.E. Bonner et al., {\it Phys. Rev. Lett.} {\bf 41}, 1200
  (1978)
\bibitem{Na73a} M.M. Nagels, Th.A. Rijken, and J.J. de Swart, {\it Ann. Phys.} 
  (NY) {\bf 79}, 338 (1973)
\bibitem{Gi} B.F. Gibson, {\it Nuclear Aspects of Few-Baryon Systems},
  {\it Proceedings of the XIVth European Conference 
  on Few-Body Problems in Physics}, Amsterdam, The Netherlands
  (B.L.G. Bakker, R. van Dantzig, eds, 1993), see also these proceedings.
\bibitem{Mae} P.M.M. Maessen, private communication
\bibitem{Al68} G. Alexander, U. Karshon, A. Shapira, G. Yekutieli,
 R. Engelmann, H. Filthuth, and W. Lughofer, {\it Phys. Rev.} {\bf 173}, 
  1452 (1968)
\bibitem{Se68} B. Sechi-Zorn, B. Kehoe, J. Twitty, and R.A. Burnstein, 
  {\it Phys. Rev.} {\bf 175}, 1735 (1968)
\bibitem{Ka71} J.A. Kadijk, G. Alexander, J.H. Chan, P. Gaposchkin, 
 and G. Trilling, {\it Nucl. Phys.} {\bf B27}, 13 (1971)
\bibitem{Ha77} J.M. Hauptman, J.A. Kadijk, and G.H. Trilling, {\it Nucl. Phys.}
  {\bf B125}, 29 (1977)
\bibitem{Ei71} F. Eisele, H. Filthuth, W. F\"olisch, V. Hepp, E. Leitner,
  and G. Zech, {\it Nucl. Phys.} {\bf B37}, 204 (1971)
\bibitem{En66} R. Engelmann, H. Filthuth, V. Hepp, and E. Kluge, {\it Phys.
  Lett.} {\bf 21}, 587 (1966)
\bibitem{He68} V. Hepp, and H. Schleich, {\it Z. Phys.} {\bf 21}, 587 (1968)
\bibitem{Ba87} P.D. Barnes et al., {\it Phys. Lett.} {\bf B 189}, 249 (1987); 
  {\bf B 199}, 147 (1987); {\bf B 229}, 432 (1989); 
  {\it Nucl. Phys.} {\bf A 526}, 575 (1991)
\bibitem{Ba90} P.D. Barnes et al., {\it Phys. Lett.} {\bf B 246}, 273 (1990)
\bibitem{Da64} R.H. Dalitz, and F. Von Hippel, {\it Phys. Lett.} {\bf 10},
  153 (1964)
\bibitem{Sw63} J.J. de Swart, {\it Rev. Mod. Phys.} {\bf 35}, 916 (1963)
\bibitem{Ko69} J.J.J. Kokkedee, {\it The Quark Model}, Frontiers in Physics,
  (W.A. Benjamin Inc., 1969)
\bibitem{Oa63} R.J. Oakes, {\it Phys. Rev.} {\bf 131}, 2239 (1963)
\bibitem{Iw64} S. Iwao, {\it Nuovo Cimento} {\bf 34}, 1167 (1964)
\bibitem{So64} P.O. deSouza, and G.A. Snow, {\it Phys. Rev.} {\bf 135 B}, 
  565 (1964)
\bibitem{Do90} C.B. Dover and H. Feshbach, {\it Ann. Phys.} (NY) {\bf 198}, 321 
  (1990)
\bibitem{Na73b} M.M. Nagels, Th.A. Rijken, and J.J. de Swart, {\it Phys. Rev.
  Lett.} {\bf 31}, 569 (1973)
\bibitem{potmod93} V.G.J. Stoks, R.A.M. Klomp, C.P.F. Terheggen, and J.J. de
  Swart, {\it Construction of high-quality $N\!N$-potential models},
  Nijmegen-report THEF-NYM-93.05, submitted for publication
\bibitem{Bo77} K. Holinde and R. Machleidt, {\it Nucl. Phys.} {\bf A 280}, 429 
  (1977)
\bibitem{Bo87} R. Machleidt, K. Holinde, and Ch.\ Elster, {\it Phys. Rep.} 
  {\bf 149}, 1 (1987)
\bibitem{BoPRC} J. Haidenbauer and K. Holinde, {\it Phys. Rev.} {\bf C 40}, 2465
  (1989)
\bibitem{BoPRep} R. Machleidt, {\it Adv. Nucl. Phys.} {\bf 19}, 189 (1989)
\bibitem{Mac71} W. Ma\^cek, M.M. Nagels, Th.A. Rijken, and J.J. de Swart 
  (unpublished, 1978)
\bibitem{St93} V. Stoks, R. Timmermans, and J.J. de Swart, {\it Phys. Rev.}
  {\bf C 47}, 512 (1993)
\bibitem{Du83} O. Dumbrajs, et al., {\it Nucl. Phys.} {\bf B 216}, 277 (1983)
\bibitem{Ti91} R.G.E. Timmermans, Th.A. Rijken, and J.J. de Swart,
  {\it Phys. Lett.} {\bf B 257}, 227 (1991)
\bibitem{Sw77} J.J. de Swart, and M.M. Nagels, {\it Fortschr. Phys.} 
  {\bf 28}, 215 (1977)
\bibitem{OZI} S. Okubo, {\it Phys. Lett.} {\bf 5}, 165 (1963) \\
  G. Zweig, CERN Report No.\ 8419/TH142 (1964, unpublished) \\
  J. Iizuka, K. Okada, and O. Shito, {\it Prog. Theor. Phys.} {\bf 35}, 1061
  (1966)
\bibitem{Sa60} J.J. Sakurai, {\it Ann. Phys.} (NY) {\bf 11}, 1 (1960)
\bibitem{Sa69} J.J. Sakurai, {\it Currents and Mesons} (University of
  Chicago Press, Chicago, 1969)
\bibitem{Ho75} G. H\"ohler and E. Pietarinen, {\it Nucl. Phys.} {\bf B 95}, 210 
  (1975);  
  R. Koch and E. Pietarinen, {\it Nucl. Phys.} {\bf A 336}, 331 (1980)
\bibitem{Sak65} B. Sakita and K.C. Wali, {\it Phys. Rev.} {\bf 139}, B1355 
  (1965)
\bibitem{Ho60} N. Hoshizaki, I.  Lin, and S. Machida, {\it Prog. Theor. Phys.}
  (Kyoto) {\bf 24}, 480 (1960);  
  N. Hoshizaki, S. Otsuki, W. Watari, and M. Yonezawa, {\it Prog. Theor. Phys.}
  (Kyoto) {\bf 27}, 1199 (1962)
\bibitem{Br64} R.A. Bryan, and B.L. Scott, {\it Phys. Rev.} {\bf 135}, B 434 
  (1964); {\it Phys. Rev.} {\bf 177}, 1435 (1969)
\bibitem{Sv92} M. Svec, A. de Lesquen, and L. van Rossum, {\it Phys. Rev.}
  {\bf D 45}, 1518 (1992)
\bibitem{Bi71} J. Binstock and R.A. Bryan, {\it Phys. Rev.} {\bf D 4}, 
  1341 (1971);
  R.A. Bryan and A. Gersten, {\it Phys. Rev.} {\bf D 6}, 341 (1972)
\bibitem{Sch71} J. Schwinger, {\it Phys. Rev.} {\bf D 3}, 1967 (1971)
\bibitem{PDG92} Particle Data Group, {\it Phys. Rev.} {\bf D 45}, S1 (1992)
\bibitem{Ae80} A.T. Aerts, P.J. Mulders, and J.J. de Swart, {\it Phys. Rev.} 
  {\bf D 21}, 1370 (1980)
\bibitem{Ja77} R.L. Jaffe, {\it Phys. Rev.} {\bf D 15}, 267 (1977);
  {\it Phys. Rev.} {\bf D 15}, 281 (1977)
\bibitem{Svec92} M. Svec, A. de Lesquen, and L. van Rossum, {\it Phys. Rev.}
  {\bf D 46}, 949 (1992)
\bibitem{Ba66} V. Barger and M. Olsson, {\it Phys. Rev.} {\bf 148}, 1428 (1966)
\bibitem{Do68} R. Dolen, D. Horn, and C. Scjhmid, {\it Phys. Rev.} {\bf 166},
 1768 (1968);  
 H. Harari, {\it Phys. Rev. Lett.} {\bf 20}, 1395 (1968)
\bibitem{Jo62} C.E. Jones, Lawrence Radiation Laboratory Report UCRL-10700
 (unpublished, 1962);  
 N.N. Khuri, {\it Phys. Rev.} {\bf 130}, 429 (1963)
\bibitem{Rij75} Th.A. Rijken, thesis University of Nijmegen (1975);  
  Th.A. Rijken, {\it Ann. Phys.} (NY) {\bf 164}, 1 and 23 (1985)
\bibitem{Lo75} F.E. Low, {\it Phys. Rev.} {\bf D 12}, 163 (1975);  
 S. Nussinov, {\it Phys. Rev. Lett.} {\bf 34}, 1286 (1975)
\bibitem{He92} T. Henkes, et al., {\it Phys. Lett.} {\bf B283}, 155 (1992);  
 A.M. Smith, et al., {\it Phys. Lett.} {\bf B163}, 267 (1985)
\bibitem{La87} P.V. Landshoff and O. Nachtmann, {\it Z. Phys. C} {\bf 35}, 405
  (1987)
\bibitem{Pu81} J. Pumplin and E. Lehman, {\it Z. Phys. C} {\bf 9}, 25 (1981);  
 J.F. Gunion and D.E. Soper, {\it Phys. Rev.} {\bf D 9}, 2617 (1977)
\bibitem{Sa51} E.E. Salpeter and H.A. Bethe, {\it Phys. Rev.} {\bf 84}, 1232
  (1951);  
  J.M. Charap and S.P. Fubini, {\it Nuovo Cimento} {\bf 14}, 540 (1959);  
 F. Gross, {\it Phys. Rev.} {\bf 186}, 1448 (1969)
\bibitem{Rij94} Th.A. Rijken, in preparation (1994).
\end{thebibliography}
\end{document}